\documentclass[a4paper,11pt]{article}
\pdfoutput=1 % if your are submitting a pdflatex (i.e. if you have
\usepackage{jinstpub} % for details on the use of the package, please
\title{\boldmath Studies of irradiated AMS H35 CMOS detectors for the ATLAS tracker upgrade}

\author[a,1]{Emanuele Cavallaro,\note{Corresponding author.}}
\author[a]{Raimon Casanova,}
\author[a]{Fabian F\"orster,}
\author[a,b]{Sebastian Grinstein,}
\author[a]{J\"orn Lange,}
\author[c]{Gregor Kramberger,}
\author[c]{Igor Mandi\'c,}
\author[a]{Carles Puigdengoles}
\author[a]{and Stefano Terzo}

\affiliation[a]{Institut de F\'{i}sica d'Altes Energies (IFAE), The Barcelona Institute of Science and Technology (BIST), 08193 Bellaterra (Barcelona), Spain}
\affiliation[b]{Instituci\'{o} Catalana de Recerca i Estudis Avan\c{c}ats (ICREA), Pg. Llu\'{i}s Companys 23, 08010 Barcelona, Spain}
\affiliation[c]{Jo\u{z}ef Stefan Institute (JSI), Ljubljana, Jamova cesta 39, 1000 Ljubljana, Slovenia}

\emailAdd{ecavallaro@ifae.es}

\abstract{

Silicon detectors based on the HV-CMOS technology are being investigated as possible candidate for the outer layers of the ATLAS pixel detector for the High Luminosity LHC.
In this framework the H35Demo ASIC has been produced in the $350 \, \mathrm{nm}$ AMS technology (H35).
The H35Demo chip has a large area ($18.49 \times 24.40 \, \mathrm{mm^2}$) and includes four different pixel matrices and three test structures.
In this paper the radiation hardness properties, in particular the evolution of the depletion region with fluence is studied using edge-TCT on test structures.
Measurements on the test structures from chips with different substrate resistivity are shown for non irradiated and irradiated devices up to a cumulative fluence of $2 \cdot 10^{15} \, \mathrm{1\,MeV\, n_{eq} / cm^{2}}$.

}


\arxivnumber{1611.04970} % only if you have one

\proceeding{Topical Workshop on Electronics for Particle Physics\\
  26-30 September 2016\\
  Karlsrhue, Germany}

\begin{document}
\maketitle
\flushbottom

\section{Introduction}
\label{sec:intro}

The High Luminosity Large Hadron Collider (HL-LHC)~\citep{HLLHC} is currently planned to start operating in 2026. The project foresees an improvement of the luminosity capability up to $7.5 \cdot 10^{34} \,\mathrm{cm^{-2}s^{-1}}$
corresponding to an average of $~200$ proton-proton interactions per bunch crossing.

The ATLAS detector will need to be upgraded to cope with the luminosity of the HL-LHC.
The new detector will need to sustain a higher trigger rate with a higher level of pile-up and the innermost layer of the pixel detector will have to maintain good performances up to fluences of  the order of $ 10^{16} \, \mathrm{ 1\,MeV \, n_{eq} / cm^{2}}$.
In order to meet these requirements the current ATLAS Inner Detector will be replaced by a full silicon detector named Inner Tracker (ITk).

The current design foresees nine silicon layers, with the inner five made of pixel detectors and the outer ones of strip detectors.
The total surface of the pixel detector is estimated to be of the order of $10 \, \mathrm{m^2}$, being the largest pixel detector ever built.
The ITk innermost layers will consist of hybrid detectors, with silicon sensors bump-bonded to front-end chips. 
This is currently the most radiation hard solution but, mostly due to bump-bonding, its production cost is high.
Thus a more cost effective solution is being investigated for the outer layers where the level of radiation is about an order of magnitude lower and the requirements can be relaxed.

High voltage CMOS (HV-CMOS) is a reliable technology commonly used in the industries but still new in the field of particle detectors which is a candidate for the outermost layer of the pixel detector. 
Such as standard CMOS and monolithic active pixel sensors (MAPS) it offers the opportunity to produce thin sensors with in-pixel analog and digital electronics. In addition it allows to apply bias voltages of the order of $100 \, \mathrm{V}$. The higher bias gives a larger depleted volume and makes the charge collection faster since its main contribution comes from the drift of the charge carriers and not from their diffusion. 
HV-CMOS technologies can be used to produce fully monolithic sensors, where no additional front-end chip is required, or capacitively coupled devices where the chip sensor is connected to a read-out chip with non conducting glue.
The opportunity of using a commercial technology and avoiding bump-bonding makes HV-CMOS devices a cost effective alternative to the standard hybrid detectors.
During the last years HV-CMOS devices from different foundries, with different technologies and on different substrate resistivities have been investigated~\citep{Peric_HVCMOS_Overview_2015,Gregor_JINST_2016,Igor_28th_RD50,Marcos_2016}. 
The results show an initial increase of the depletion depth after irradiation described by an effective reduction of the acceptor concentration in the substrate. This phenomenon, known as acceptor removal, was found to depend on the initial acceptor concentration and thus on the substrate resistivity.

In this paper the evolution of the depletion depth with irradiation on sensors produced in the AMS H35 technology~\citep{AMS} with silicon wafers of different resistivities is presented. All the sensors included in this study present the same test structure allowing a direct comparison of the results.
An irradiation campaign up to $2 \cdot 10^{15} \, \mathrm{ 1\,MeV \, n_{eq} / cm^{2}}$ has been carried out to investigate radiation damages equivalent to the ones expected after 10 years of operation at the HL-LHC in the outer pixel layer.

In section~\ref{sec:chip} a description of the chip and its test structure is given. Section~\ref{sec:TCT_setup} describes the experimental set-up and the measurements before and after the samples irradiation are discussed in sections~\ref{sec:measurements} and~\ref{sec:irradiation} respectively. Conclusions are presented in section~\ref{sec:conclusions}.

\section{The H35Demo chip}
\label{sec:chip}
The H35Demo ASIC~\citep{Eva_H35Demo} is a demonstrator chip produced in the $350 \, \mathrm{nm}$ AMS technology by the collaboration of the Karlsruhe Institute of Technology (KIT), the Institut de F\'isica d'Altes Energies of Barcelona (IFAE), the University of Geneva and the University of Liverpool.
It is a large area chip, $18.49 \times 24.40 \, \mathrm{mm^2}$, developed to investigate the possibility to install this technology in the ATLAS ITk.
The usual substrate wafer resistivity for high energy physics application is of the order of the $\mathrm{k\Omega cm}$ while the standard value for the AMS H35 is $20 \, \mathrm{\Omega cm}$. Nevertheless the H35Demo chip has been produced in addition also on wafers of non standard resistivities: $80 \, \mathrm{\Omega cm}$, $200 \, \mathrm{\Omega cm}$ and $1 \, \mathrm{k \Omega cm}$.

\subsection{Pixel Matrices}
The H35Demo chip contains four pixel matrices: a standalone nMOS matrix, two analog matrices and a standalone CMOS matrix. The pixel size is $50 \times 250 \, \mathrm{\mu m^2}$ in all the matrices while the number of pixels is $16 \, \mathrm{rows} \times 300 \, \mathrm{columns} $ in the nMOS and CMOS matrices and $23 \, \mathrm{rows} \times 300 \, \mathrm{columns} $ in the analog matrices.
The sensors are implemented through the p-n junctions made by the deep N wells in the p-doped substrate.
 
\begin{figure}[htb]
\begin{center}
  \includegraphics[width=0.32\columnwidth]{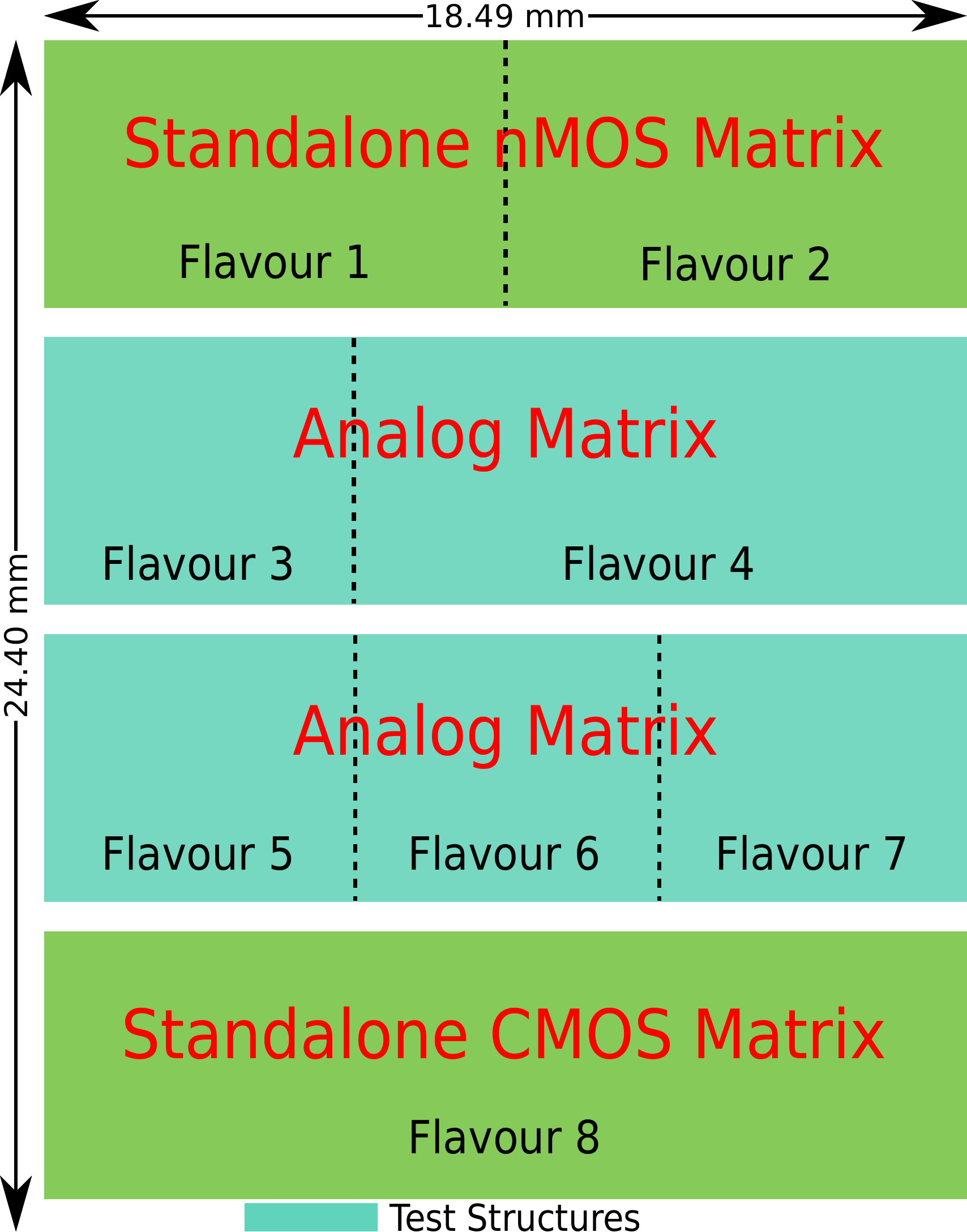}
  \caption{Layout of the H35Demo chip showing the four pixel matrices with their flavours and the position of the test structures.}
\label{fig:H35Demo}
\end{center}
\end{figure}

Different pixel flavours are included in the matrices, each one with different characteristics in terms of in-pixel electronics: in the standalone nMOS matrix half of the pixels have time-walk compensation, the analog matrices have five total pixel flavours for different amplification gains and speeds, the CMOS matrix contains only one pixel flavour, see figure~\ref{fig:H35Demo}.

Each pixel of the matrices has an output attached to a pad for interconnection to a FE-I4 chip~\citep{FEI4_NIM}. 
Both analog and standalone matrices can be read out through the FE-I4 chip allowing to test the performance of the standalone matrices in the monolithic and capacitively coupled configurations.

\subsection{H35Demo test structure}

In addition to these four matrices the chip also contains three test structures. Two $3 \times 3$ matrices that differ in the pixel size and in the presence of in-pixel amplification, and a test structure for capacitance measurement.

\begin{figure}[htb]
\begin{center}
		\includegraphics[width=0.6\columnwidth]{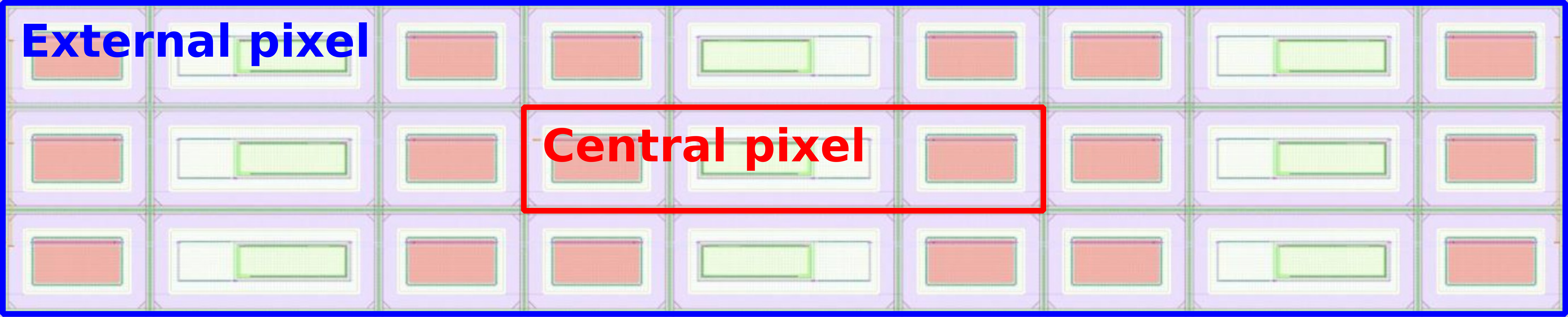}
		\vspace{0.05\columnwidth}
		
 		\includegraphics[width=0.6\columnwidth]{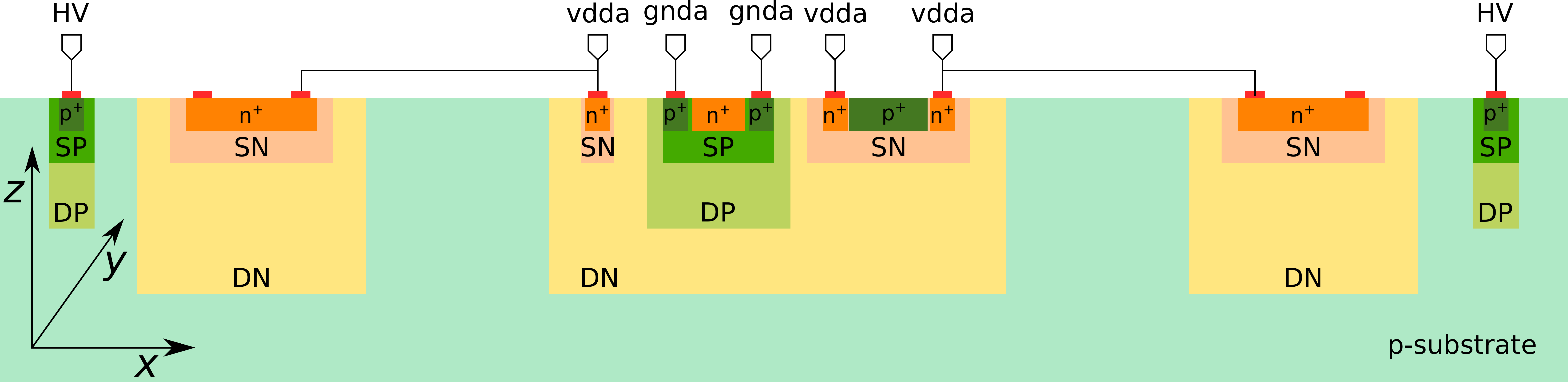}
	\caption{H35Demo test structure layout with the central pixel marked in red (top) and cross section of a pixel (bottom). DN(P) and SN(P) denote the deep and shallow N(P) wells.}
	\label{fig:test_structure}
\end{center}
\end{figure}

The subject of this study is a test structure made of a $3 \times 3$ pixel matrix. Each pixel of $50 \times 250 \, \mathrm{\mu m^2}$ contains three deep N wells: a central N well of $50 \times 110 \, \mathrm{\mu m^2}$ containing a deep P well, and two external wells of $50 \times 70 \, \mathrm{\mu m^2}$, see figure~\ref{fig:test_structure}.
The pixels of this test structure are identical to those of the matrices except that the in-pixel electronics are not present. 
The signal of the central pixel (marked in red in figure~\ref{fig:test_structure}) and of the surrounding eight pixels, shorted together, can be read out separately.

The devices tested for this study are three samples from $700 \, \mathrm{\mu m}$ thick wafers of different nominal resistivities: $80 \, \mathrm{\Omega cm}$ (Sensor 1), $200 \, \mathrm{\Omega cm}$ (Sensor 2) and $1 \, \mathrm{k\Omega cm}$ (Sensor 3).
The test structures are biased through a bias grid on the top surface running along the pixel perimeter, see figure~\ref{fig:test_structure}.
For all the plots shown in this paper the coordinate reference system shown in figure~\ref{fig:test_structure} is used.

\section{TCT Set-up}
\label{sec:TCT_setup}
The purpose of the measurements carried out is to measure the charge collected at different distances from the implant surface.
To perform these measurements a scanning Transient-Current-Technique (TCT) set-up from Particulars has been used~\citep{particulars}.
The set-up consists of a $1064 \,\mathrm{nm}$ infra-red (IR) pulsed laser that illuminates the detector under test (DUT) which is mounted on a movable stage.
The IR light penetrates the silicon and generates electron-hole pairs along its trajectory. 
Since the attenuation length in silicon of IR light is larger than the pixel size, the generation of electron-hole pairs can be considered uniform along the pixel. The laser intensity has to be low enough to avoid plasma effects that can modify the electric field in the bulk.
The laser spot at its focus has a $\sigma \sim 8 \, \mathrm{\mu m}$. The optical system can be moved along the laser beam axis to focus the laser on the DUT.

The charge generated by the laser moves under the effect of the electric fields and induces a current pulse (waveform) that is acquired.
The purpose of the TCT is to measure the current waveform induced by the charge carriers generated by a light pulse.
The duration and repetition rate of the laser pulse are adjustable and their values have been set respectively to $500 \,\mathrm{ps}$ and $1 \, \mathrm{kHz}$. The usual step size of the movable stages is $1 \, \mathrm{\mu m}$.

The current waveform is amplified with a $53 \, \mathrm{dB}$ broadband amplifier and read out through a DRS4 
evaluation board with $700 \, \mathrm{MHz}$ of bandwidth and $5 \,\mathrm{GSPS}$ of sampling rate.
The DRS4 has four input channels, two of them are used to read out the waveforms from the central and the outer pixels of the test structure, another reads out the signal form the laser driver used to trigger the data acquisition and the fourth channel acquires the signal from the beam monitoring system.
This last signal is generated splitting the laser pulse to illuminate a monitoring diode in addition to the DUT. It is important to monitor the laser intensity that, during long scans, can vary significantly due to changes in environmental conditions such as the room temperature.

\subsection{Edge TCT}

In order to study the evolution of the depletion depth the sensors are mounted on a special holder that allows to point the laser on the sensor edge rather than its top surface, in a configuration called edge-TCT~\citep{Gergor_TCT_2010}.
The edge surface is polished to avoid undesired scattering of the laser light entering into the sensor substrate. The edge is first smoothed with the use of a polishing sheet with diamond grains of $3 \, \mathrm{\mu m}$ and then with a diamond paste of $1-10 \, \mathrm{\mu m}$ grain size.

\section{Depletion depth measurement}
\label{sec:measurements}
A MIP crossing a silicon sensor perpendicular  to the surface induces a signal whose amplitude is proportional to the depth of the depleted volume. 
For this reason it is important to know the evolution of the depletion depth with voltage and irradiation.
To measure the voltage dependence of the depletion depth a scan along the thickness of each sensor is performed at increasing values of the bias voltage from $0 \, \mathrm{V}$ to the maximum applicable one in steps of $10 \, \mathrm{V}$. The typical breakdown voltage of the H35Demo chip is between  $100 \, \mathrm{V}$ and  $200 \, \mathrm{V}$ before irradiation.

\begin{figure}[htb]
\begin{center}
		\includegraphics[width=0.45\columnwidth]{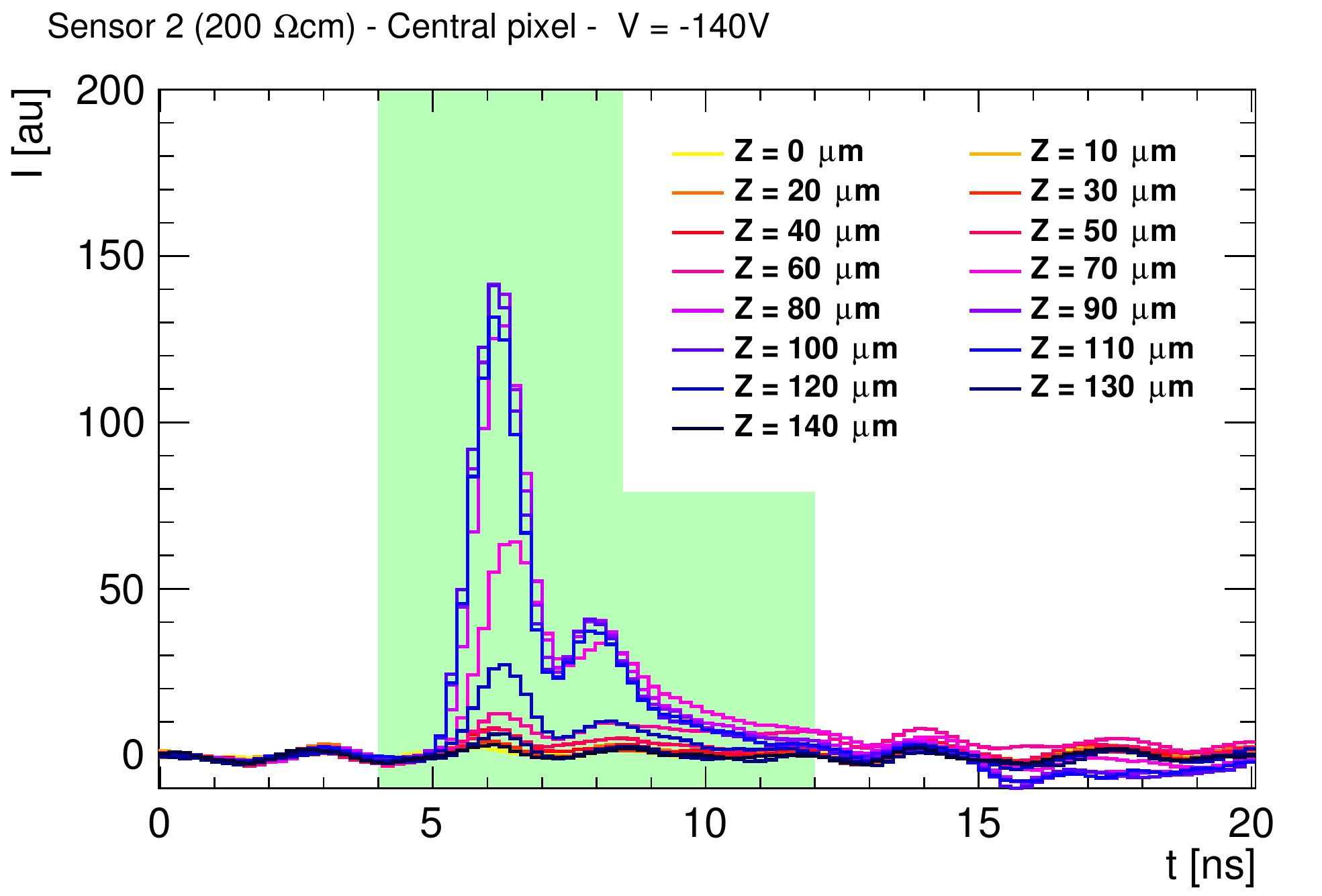}
	\caption{Current waveforms corresponding to different illumination depths on sample 2 at $140 \, \mathrm{V}$ bias voltage. The green shadow indicates the integration range to compute the charge.}
	\label{fig:wfm}
\end{center}
\end{figure}

At each scanning point $100$ waveforms are sampled and averaged to reduce fluctuations due to random noise.
The charge is taken as the integral of the waveform in a time window of $8 \, \mathrm{ns}$. Examples of waveforms are shown in figure~\ref{fig:wfm}.

\begin{figure}[htb]
\begin{center}
		\includegraphics[width=0.32\columnwidth]{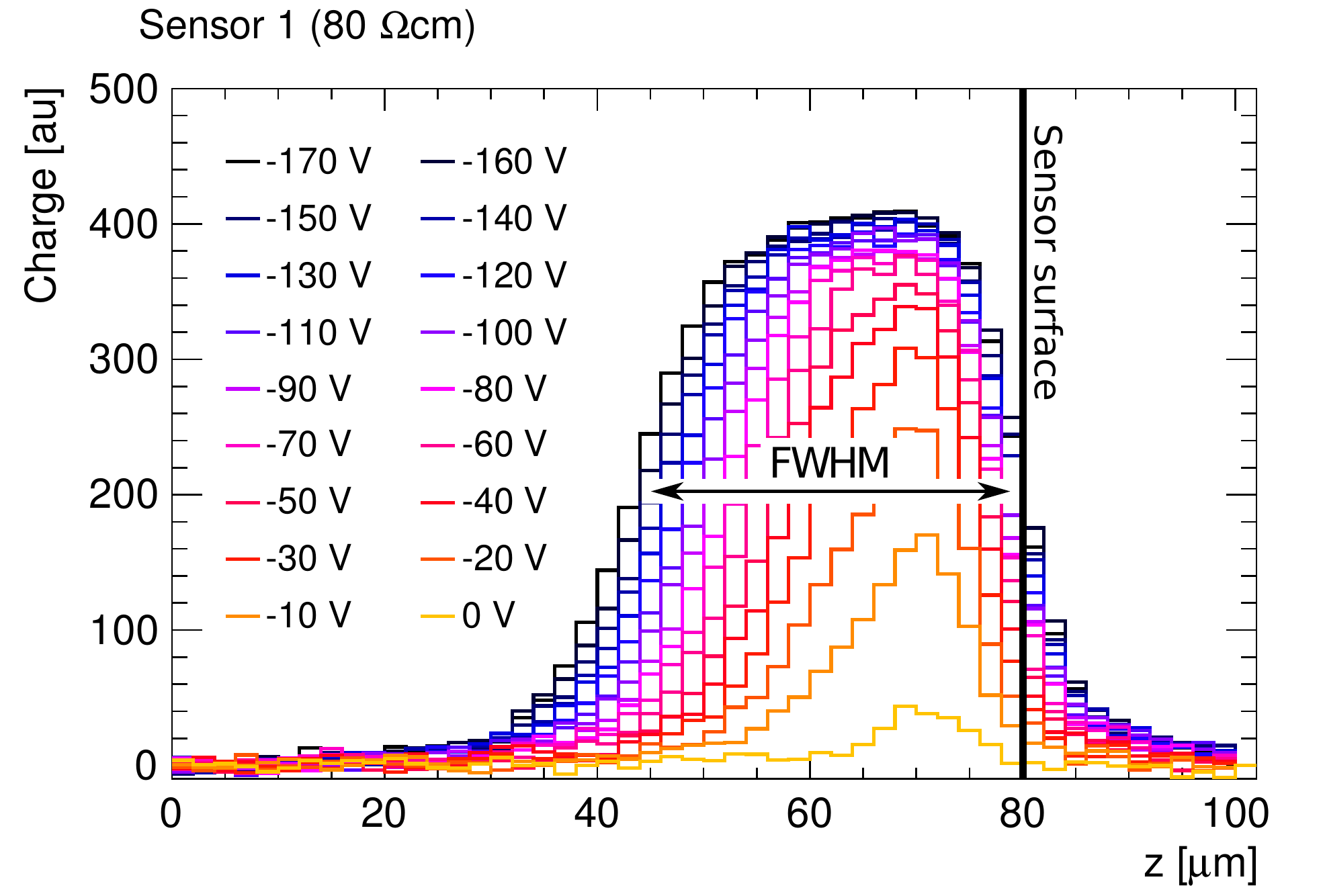}
 		\includegraphics[width=0.32\columnwidth]{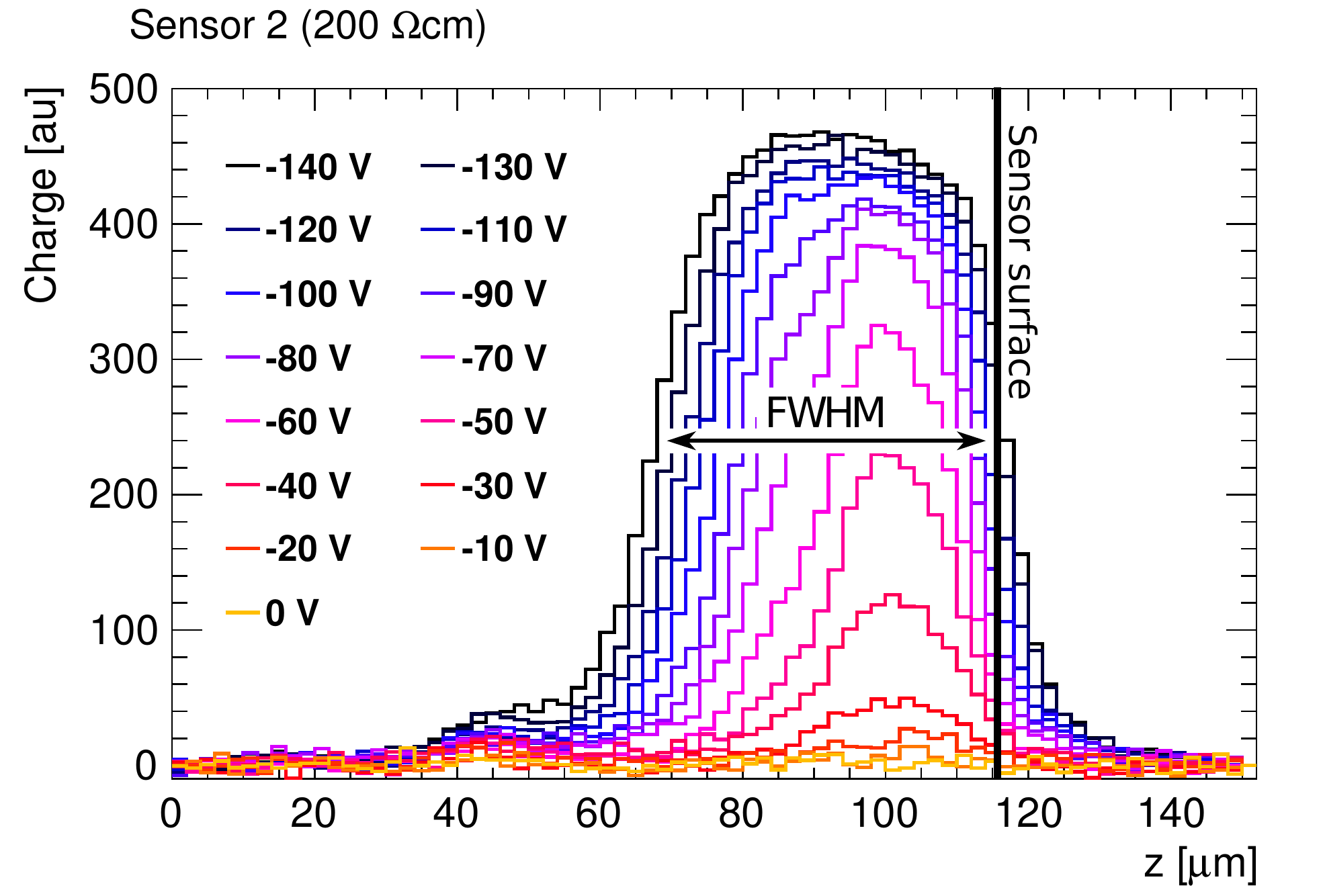}
 		\includegraphics[width=0.32\columnwidth]{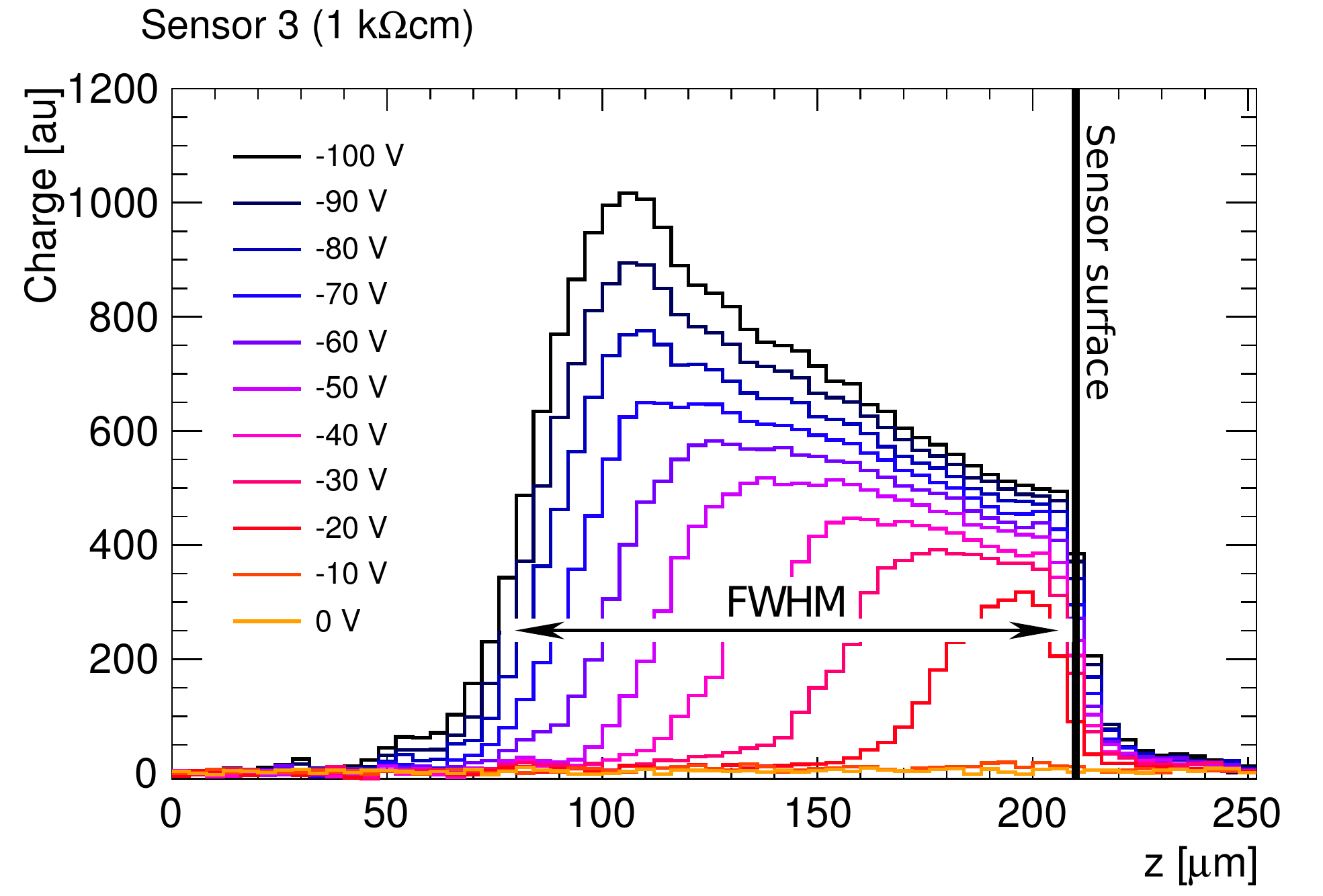}    	
	\caption{Charge collection profiles for the central pixel of the three sensors under study. The vertical lines indicate the position of the sensors top surface, the sensors extend to the left of these lines. The full width at half maximum for the largest applied bias on each sensor is shown. Note the different scales on the x axis.}
	\label{fig:profiles}
\end{center}
\end{figure}

By scanning each sensor through its thickness it is possible to obtain the charge collection profiles, shown in figure~\ref{fig:profiles}.
The plots show how by increasing the bias voltage the charge is collected to larger depth in the sensors as expected.

\begin{figure}[htb]
\begin{center}
	\includegraphics[width=0.5\columnwidth]{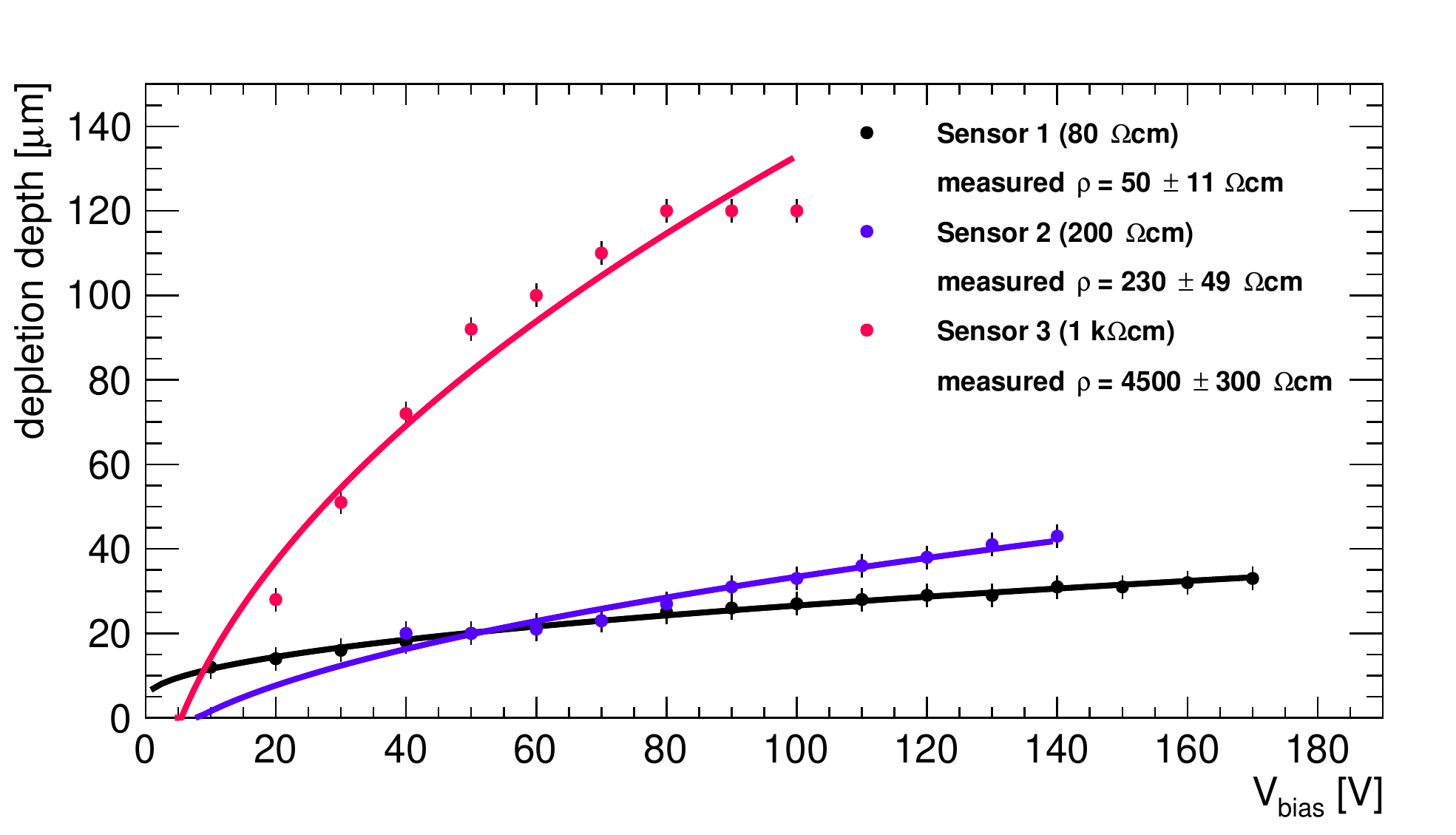}
	\caption{Depletion depth against bias voltage for the three samples with different bulk resistivities.}
\label{fig:d_vs_V}
\end{center}
\end{figure}

The charge collection profiles of sensors 1 and 2, for high enough bias voltage, show a plateau from the sensor surface to the end of the depletion depth pointing out a good charge collection uniformity along the whole space charge region. The left and right edges are smeared out mainly due to the finite laser width. The profile of sensor 3 shows instead that the charge collected by the central pixel increases with the depth until the end of the depleted volume is reached. 
A possible source of this effect can be the front biasing, comparison with simulations and with back side biased sensors are needed for a better understanding.
However this phenomenon still allows to  perform the depletion depth measurement that is taken as the full width at half maximum (FWHM) of the charge collection profiles. For sensor 3 the maximum on the sensor surface side is taken as reference for the FWHM computation.
The depletion depth measured on the three DUTs for different bias voltage are shown in figure~\ref{fig:d_vs_V}.

\begin{figure}[htb]
\begin{center}
		\includegraphics[width=0.3\columnwidth]{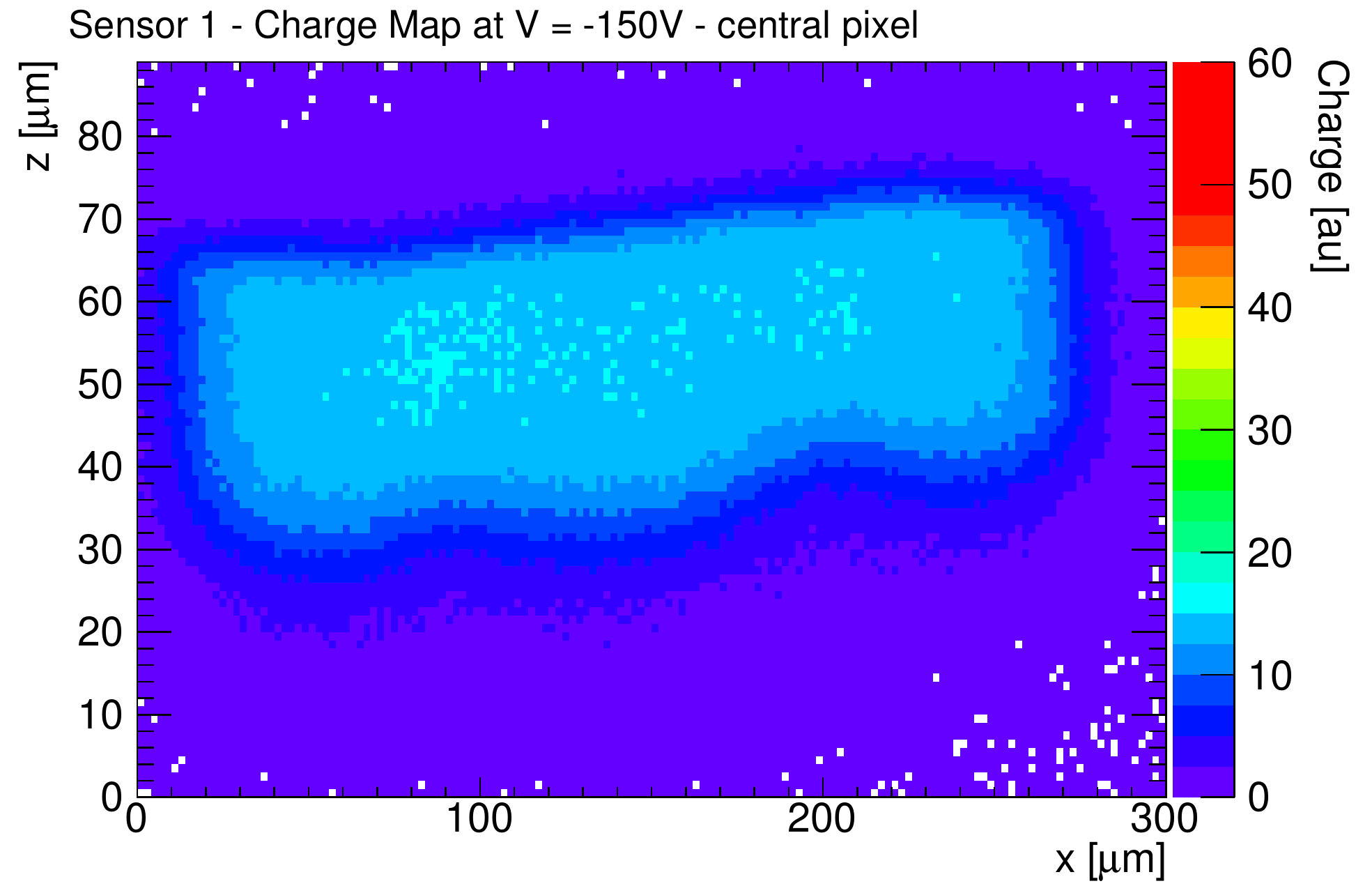}
		\includegraphics[width=0.3\columnwidth]{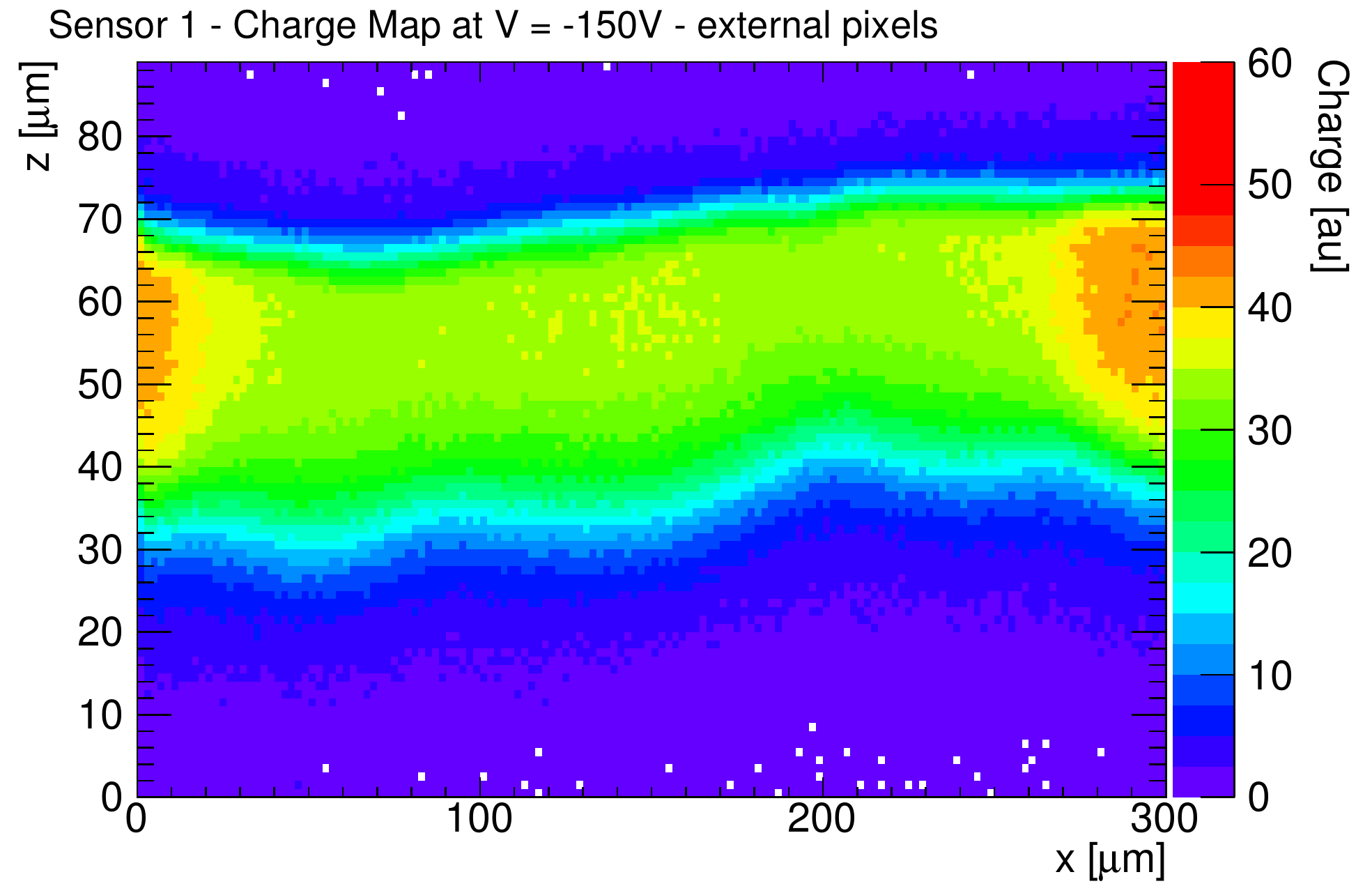}
 		\includegraphics[width=0.3\columnwidth]{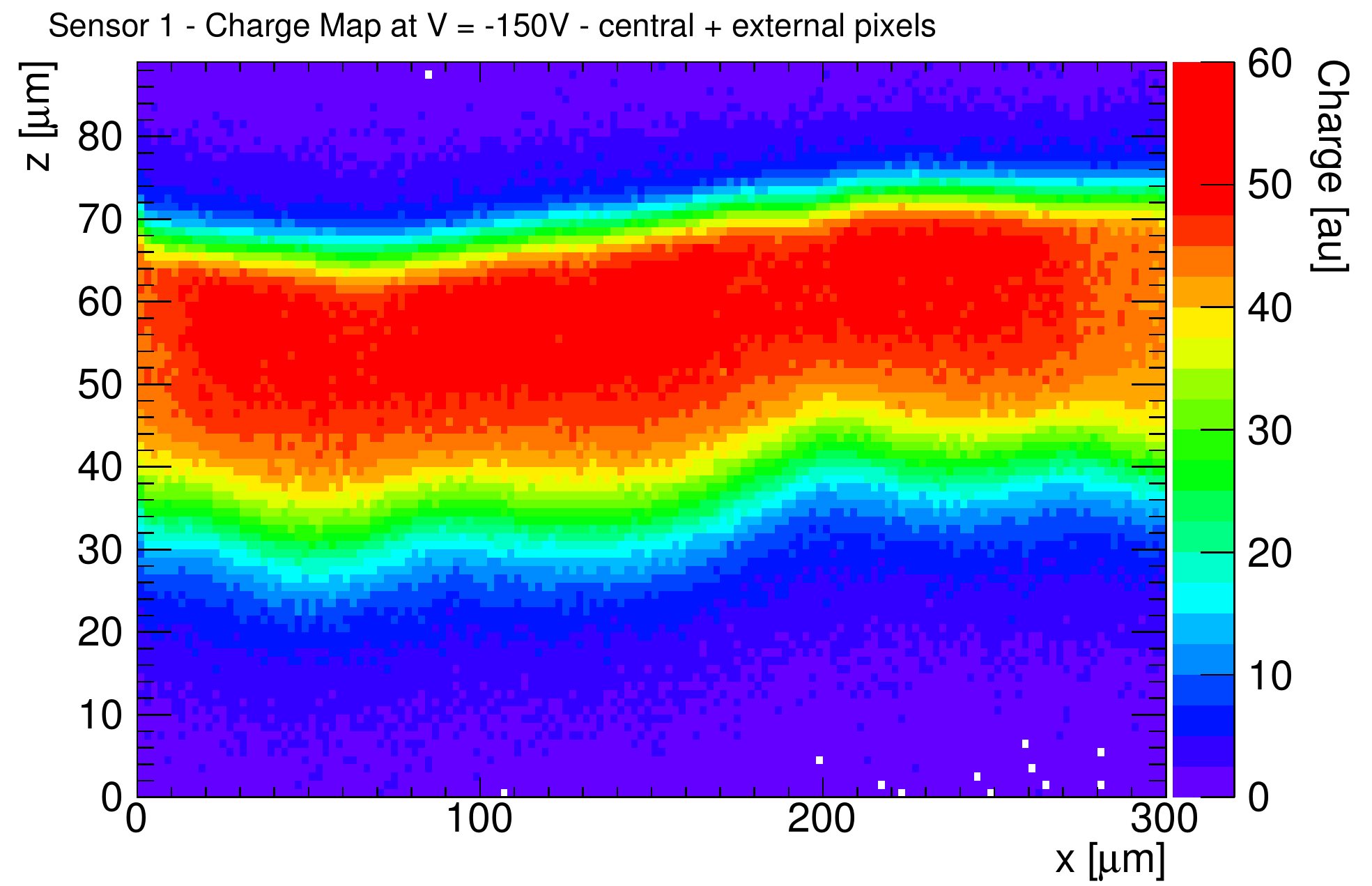}    	
		\includegraphics[width=0.3\columnwidth]{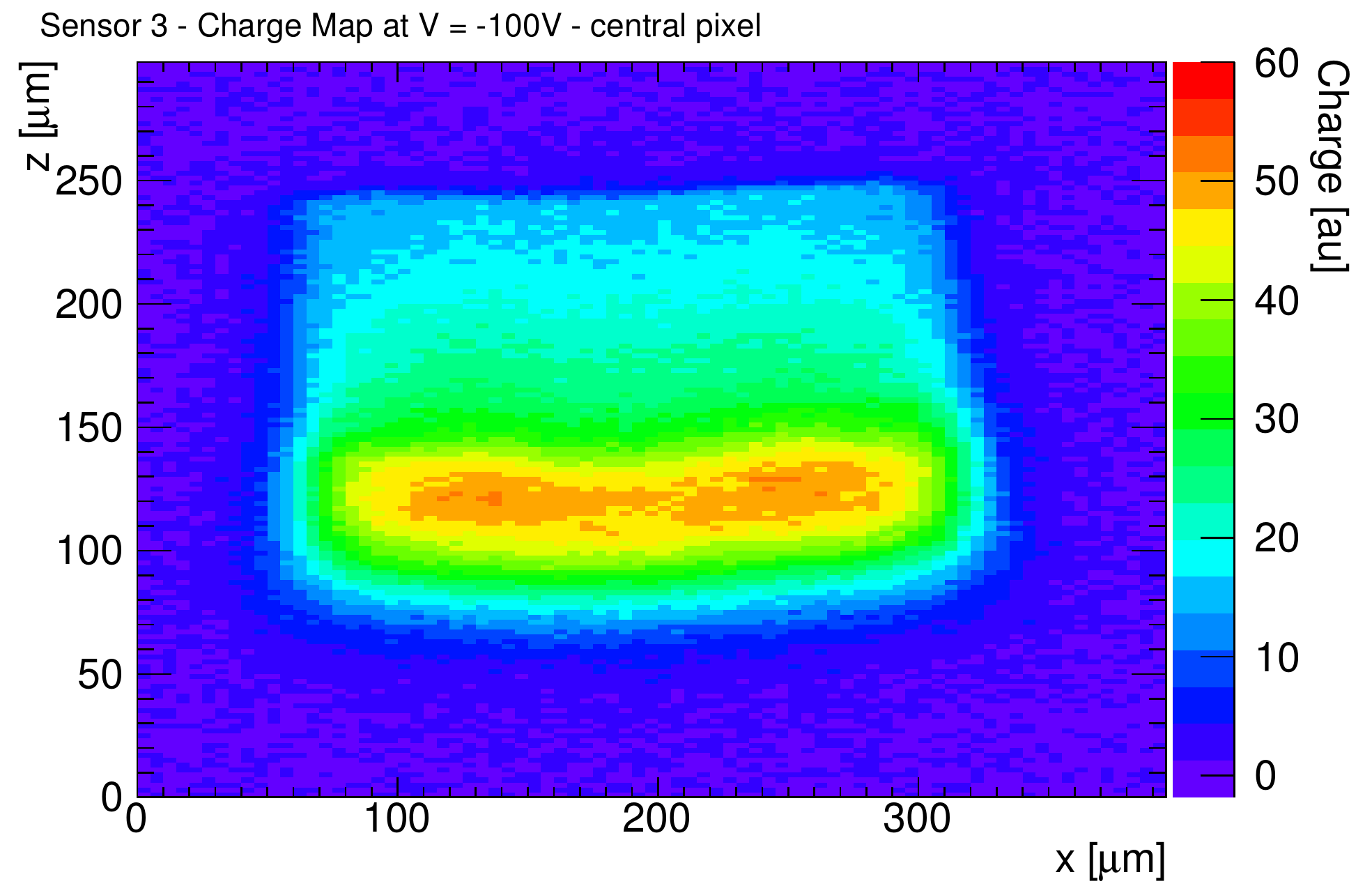}
		\includegraphics[width=0.3\columnwidth]{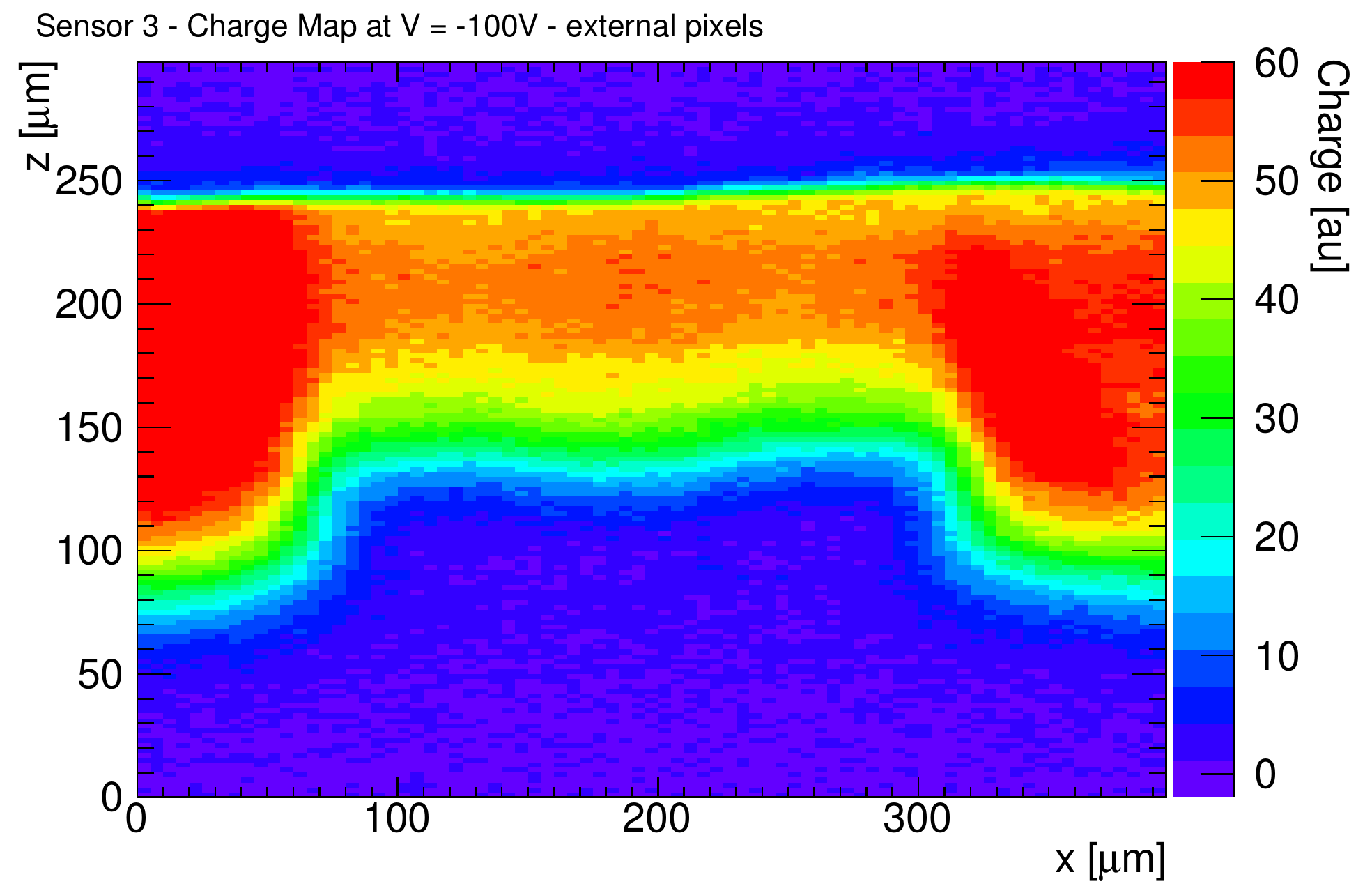}
 		\includegraphics[width=0.3\columnwidth]{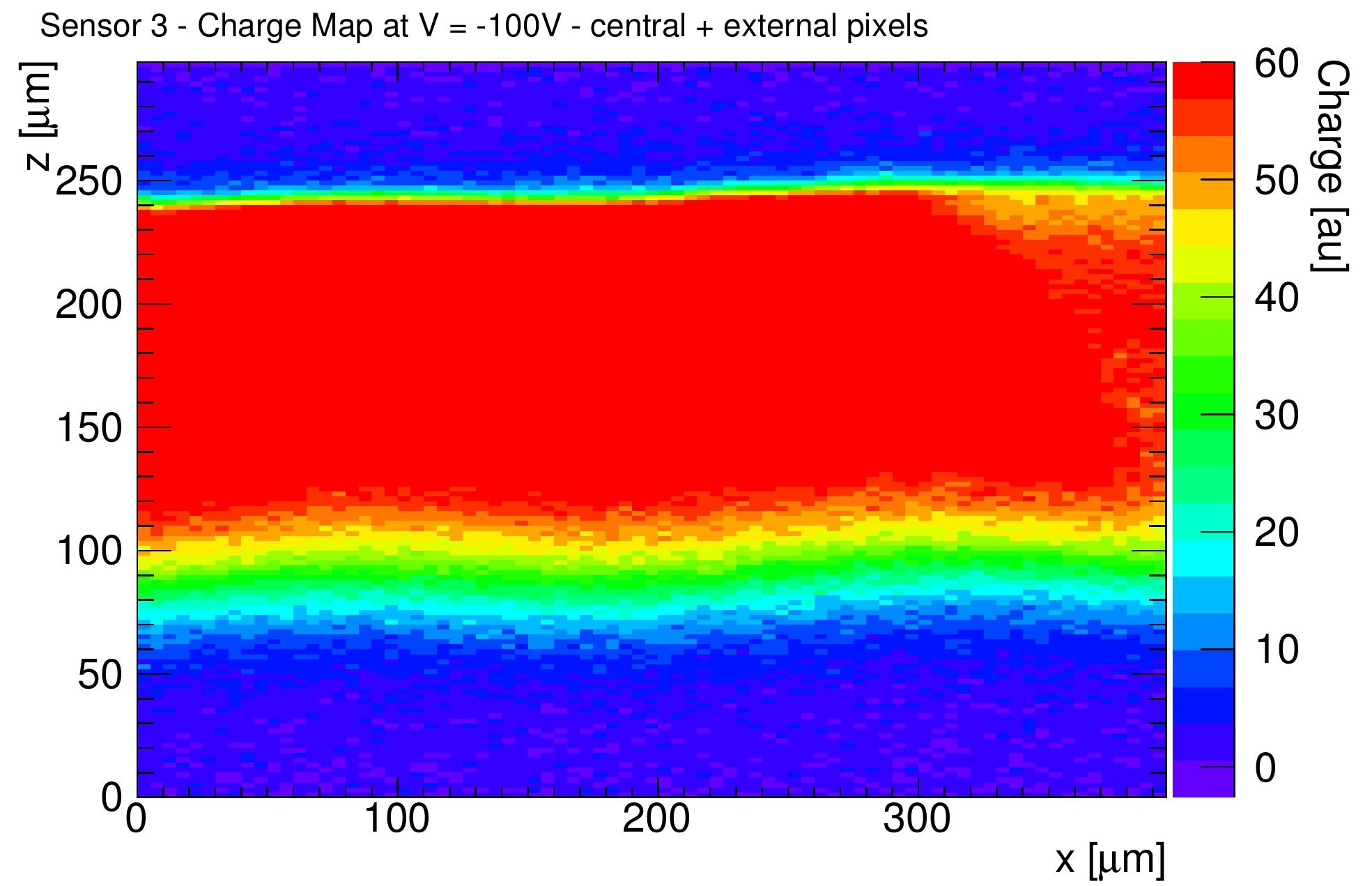}    	
	\caption{Charge collection maps for Sensor 1 (top) and Sensor 3 (bottom) respectively at a bias voltage of $150 \,  \mathrm{V}$ and $100 \,  \mathrm{V}$. From left to right, the charge collected by the central pixel, the external pixels and the sum of central and external pixels.}
	\label{fig:2Dmap}
\end{center}
\end{figure}

The charge collection maps for sensors 1 and 3 are shown in figure~\ref{fig:2Dmap} pointing out the uniformity of charge collection along the x coordinate.
There is no significant gap between the N wells within the pixel or between the pixels.
It is also visible that the increase of collected charge with depth in sensor 3 is uniform along x and corresponds to a reduction of the charge collected by its neighbours on the laser trajectory meaning that the charge carriers are collected by the central pixel although they where generated underneath its neighbours.

The relation between the depth of the depleted volume and the bias voltage applied to a diode is given by the equation:
\begin{equation}
\label{eq:d_vs_V}
d = d_0 + \sqrt{2 \varepsilon \varepsilon_0 \mu \rho V}
\end{equation}
where $d_0$ is the effective depletion depth at $V = 0 \, \mathrm{V}$, $\varepsilon \varepsilon_0$ is the silicon permittivity, $\mu$ the electron mobility in silicon and $\rho$ the substrate resistivity.
The parameter $d_0$ takes into account the built-in voltage and the finite laser size both giving a positive contribution. 

Equation~\ref{eq:d_vs_V} assumes that the bias voltage is applied from the back while in the case of the sensors used for this study the bias is applied from the top. Despite the approximation given by this model, the measured resistivity obtained by fitting the depletion depth against the bias voltage agrees with its nominal value for samples 1 and 2 where the fits return a value of $50 \pm 11 \, \mathrm{\Omega cm}$ and $230 \pm 49 \, \mathrm{\Omega cm}$ respectively. The measured resistivity of sample 3 is $4500 \pm 300 \, \mathrm{\Omega cm}$, significantly larger than the nominal value of $1 \, \mathrm{k\Omega cm}$.

The fits in figure~\ref{fig:d_vs_V} shows a negative value of $d_0$ for sensors 2 and 3 as if a turn-on threshold voltage has to be reached to make the sample able to collect charge. This effect is still under investigation, a better resolution is needed to study the charge collection profile at low voltages.

\section{Irradiation campaign}
\label{sec:irradiation}
\begin{figure}[hbt]
\begin{center}
	\includegraphics[width=0.48\columnwidth]{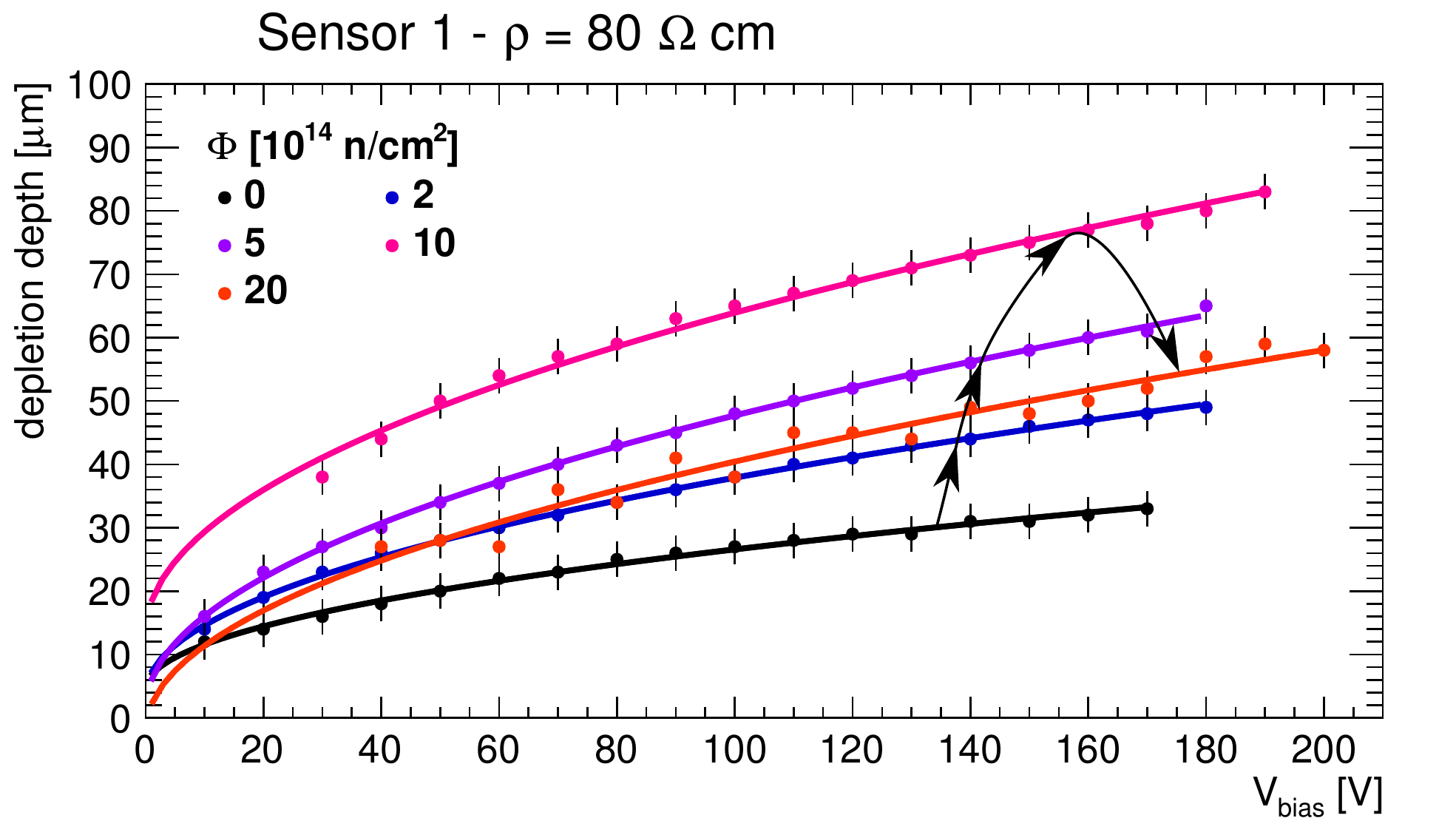}
	\includegraphics[width=0.48\columnwidth]{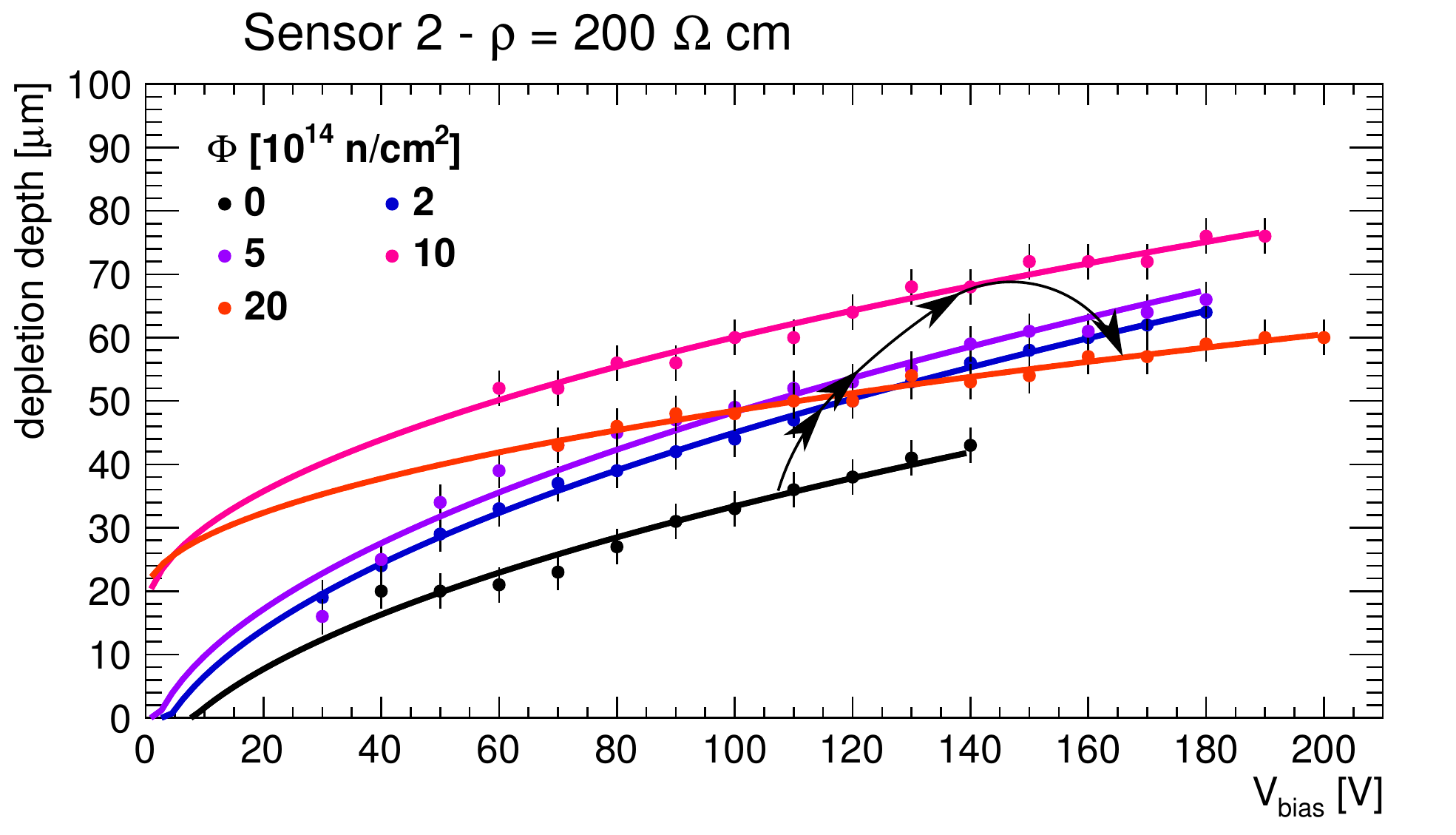}	
	\includegraphics[width=0.48\columnwidth]{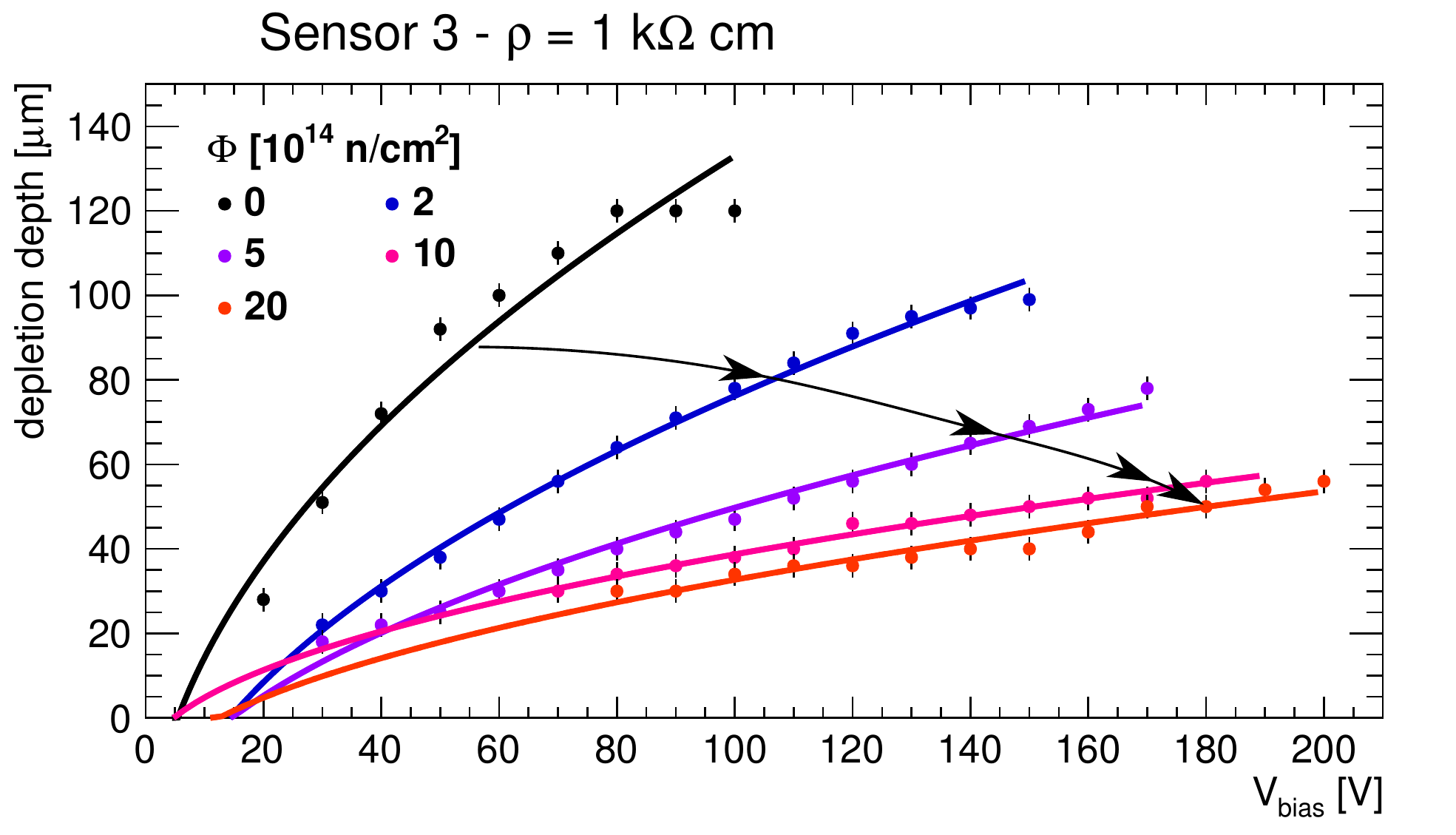}
	\caption{Depletion depth against bias voltage for the three samples for different fluence steps up to $2 \cdot 10^{15} \, \mathrm{ 1\,MeV \, n_{eq} / cm^{2}}$.  The arrows point the consecutive fluence steps.}
\label{fig:d_vs_V_irr}
\end{center}
\end{figure}

An irradiation campaign has been carried out at the TRIGA reactor of the Jo\v{z}ef Stefan Institute in Ljubljana~\citep{TRIGA}. 
The sensors studied before irradiation have been irradiated up to a fluence $\Phi$ of $2 \cdot 10^{15} \, \mathrm{ 1\,MeV \, n_{eq} / cm^{2}}$ with the following intermediate steps: $2 \cdot 10^{14} \, \mathrm{ 1\,MeV \, n_{eq} / cm^{2}}$, $5 \cdot 10^{14} \, \mathrm{ 1\,MeV \, n_{eq} / cm^{2}}$ and $10^{15} \, \mathrm{ 1\,MeV \, n_{eq} / cm^{2}}$.
After each irradiation the depletion depth as a function of the bias voltage has been measured with edge-TCT, the results are shown in figure~\ref{fig:d_vs_V_irr}.

In sensors 1 and 2 the depth of the depleted volume initially increases with irradiation reaching a maximum after a cumulated fluence of $10^{15} \, \mathrm{ 1\,MeV \, n_{eq} / cm^{2}}$ and decreases after exposition to a total fluence of $2 \cdot 10^{15} \, \mathrm{ 1\,MeV \, n_{eq} / cm^{2}}$ remaining still larger than the depletion depth before irradiation.
On the contrary sensor 3 show a reduction of the depletion depth since the first step of irradiation at $2 \cdot 10^{14} \, \mathrm{ 1\,MeV \, n_{eq} / cm^{2}}$. A depletion depth larger than $30 \, \mathrm{\mu m}$ can be achieved at any of the tested fluences on all the tested devices.
After irradiation to $2 \cdot 10^{15} \, \mathrm{ 1\,MeV \, n_{eq} / cm^{2}}$ a depletion depth larger than $50 \, \mathrm{\mu m}$ can be reached. 

The effect radiation has on the depletion depth can be explained as a radiation induced change in the effective substrate resistivity. 
Writing the resistivity in terms of the effective doping concentration $N_{eff}$, using the relation $\rho = 1/\mu e N_{eff}$ where $e$ is the elementary electric charge, equation~\ref{eq:d_vs_V} becomes:

\begin{equation}
d = d_0 + \sqrt{\frac{2\varepsilon \varepsilon_0}{e N_{eff}} V}.
\label{eq:d_vs_V_Neff}
\end{equation}
The evolution of $N_{eff}$ with fluence $\Phi_{eq}$ is described by 
\begin{equation}
N_{eff} = N_{eff0} - N_c \cdot (1 - \exp(-c \cdot \Phi_{eq})) + g_c \cdot \Phi_{eq} 
\label{eq:Neff_vs_Phi}
\end{equation}
where the $N_{eff0}$ is the initial doping concentration, $N_c$ and $c$ describes the size and the speed of the acceptor removal effect and the $g_c$ describes the radiation induced acceptor introduction~\citep{Hamburg_model}.

\begin{figure}[htb]
\begin{center}
	\includegraphics[width=0.5\columnwidth]{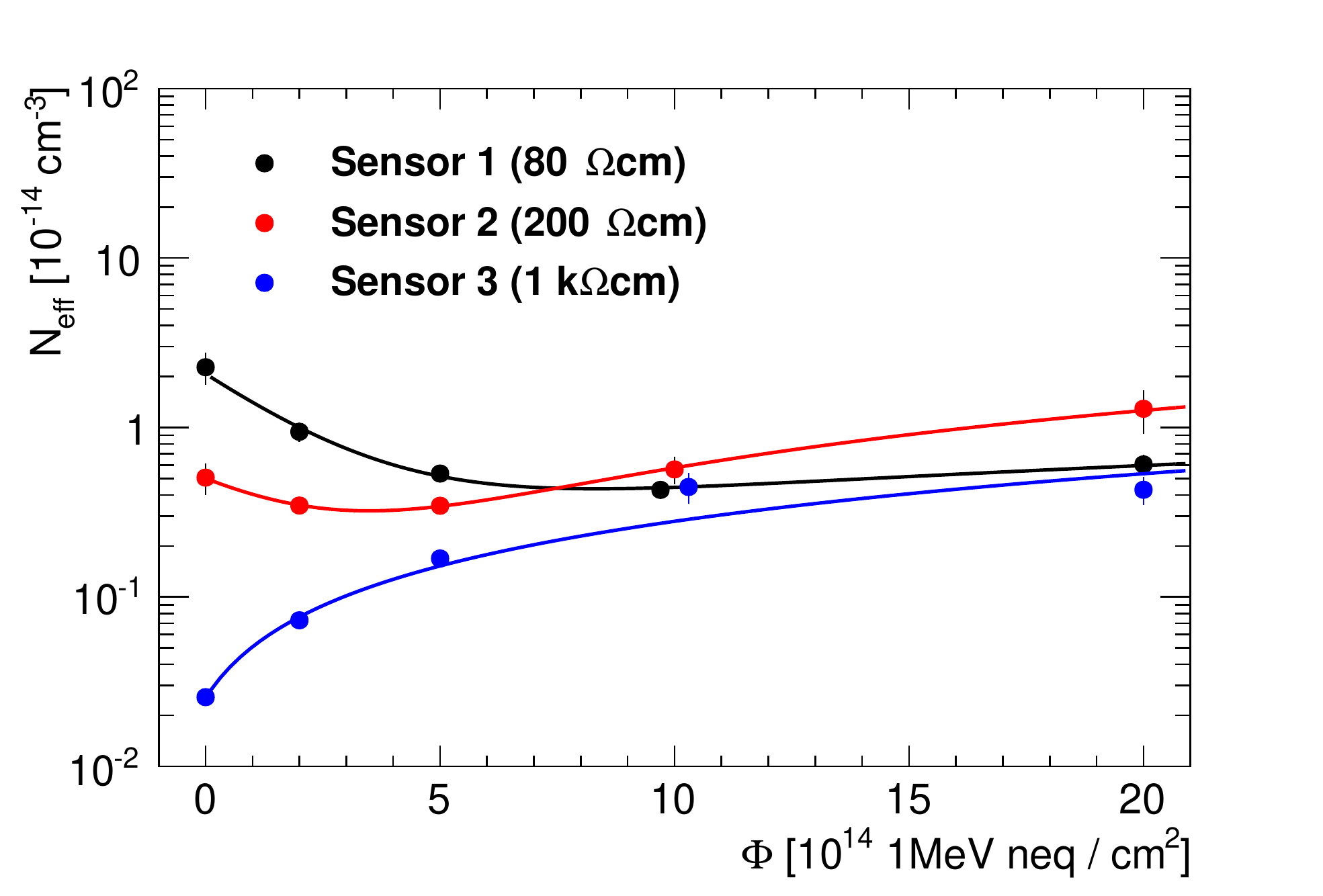}
	\caption{Measured effective dopant concentration on the samples under study at each of the tested fluences. The points at $10 \cdot 10^{14} \, \mathrm{ 1\,MeV \, n_{eq} / cm^{2}}$ of sensors 1 and 3 have been slightly shifted horizontally for better visibility.}
\label{fig:N_eff_vs_phi}
\end{center}
\end{figure}

Equation~\ref{eq:d_vs_V_Neff} can be used to obtain the value of $N_{eff}$ for each sensor and fluence by fitting the plots in figure~\ref{fig:d_vs_V_irr} and equation~\ref{eq:Neff_vs_Phi} can subsequently be used to fit $N_{eff}$ against the fluence, see figure~\ref{fig:N_eff_vs_phi} and table~\ref{tab:N_eff_vs_phi}.

\begin{table}[htbp]
\centering
\smallskip
\begin{tabular}{|r|ccc|}
\hline
&Sensor 1&Sensor 2&Sensor 3\\
\hline
$N_c \, [10^{-14} \, \mathrm{cm^{-3}}] $& $1.8\pm 0.4$ & $0.7\pm 0.5 $ & --\\
$c \, [10^{14} \, \mathrm{cm^2}]$& $0.5 \pm 0.2$ & $0.3\pm 0.3$ & --\\
$g_c \, [\mathrm{cm^{-1}}]$ & $0.02\pm 0.01$ & $0.07\pm 0.04$ & $0.03\pm  0.02$\\
\hline
\end{tabular}
\caption{\label{tab:N_eff_vs_phi} Fitted values of equation~\ref{eq:Neff_vs_Phi} parameters.}
\end{table}

Due to the low value of $N_{eff0}$ in sensor 3 the contribution of the acceptor removal term is negligible for the fluence range studied while for sensors 1 and 2 it dominates until a fluence of about $1 \cdot 10^{15} \, \mathrm{ 1\,MeV \, n_{eq} / cm^{2}}$.

The parameters of function~\ref{eq:Neff_vs_Phi} shown in table~\ref{tab:N_eff_vs_phi} match with previous results~\citep{Gregor_JINST_2016,Igor_28th_RD50} obtained for devices of $10 \, \mathrm{\Omega cm}$, $20 \, \mathrm{\Omega cm}$ and $100 \, \mathrm{\Omega cm}$. Measurements on back biased thin devices could reduce the uncertainties on the values of table~\ref{tab:N_eff_vs_phi} since all the assumptions of the model described by equation~\ref{eq:d_vs_V} would apply.

\section{Conclusions}
\label{sec:conclusions}
The depletion depth of the  H35Demo chip was studied before and after irradiation up to a fluence of  $2 \cdot 10^{15} \, \mathrm{ 1\,MeV \, n_{eq} / cm^{2}}$ on test structures on chips from different substrate resistivities: $80 \, \mathrm{\Omega cm}$, $200 \, \mathrm{\Omega cm}$ and $1 \, \mathrm{k\Omega cm}$.
Before irradiation the measured resistivities match with the nominal ones for the sensors of lower resistivity while the measured resistivity of the $1 \, \mathrm{k\Omega cm}$ sample resulted to be about four times larger.
The measurements carried out during the  irradiation campaign show that the space charge region of sensors with substrate resistivity of $80$ and $200 \, \mathrm{\Omega cm}$ increases with irradiation up to a fluence of $1 \cdot 10^{15} \, \mathrm{ 1\,MeV \, n_{eq} / cm^{2}}$ and decreases for higher fluences. 
Up to a fluence of $2 \cdot 10^{15} \, \mathrm{ 1\,MeV \, n_{eq} / cm^{2}}$ the depletion depth is always larger than the one obtained before irradiation.
The sensor with a substrate resistivity of $1 \, \mathrm{k\Omega cm}$ instead shows a notable reduction of the space charge region since the first irradiation step at $2 \cdot 10^{14} \, \mathrm{ 1\,MeV \, n_{eq} / cm^{2}}$. Nevertheless it is possible to achieve a depletion depth bigger than $30 \, \mathrm{\mu m}$ on all the tested devices at any of the tested fluences. This result does not point out any preferred or rejected resistivity value.
Work is on-going to readout the pixel matrices in order to test the H35Demo chip performance in beam tests.

Although the outer layers of the ATLAS pixel tracker will not exceed a fluence of $2 \cdot 10^{15} \, \mathrm{ 1\,MeV \, n_{eq} / cm^{2}}$ it is interesting to probe the limits of this technology. Therefore the irradiation campaign will continue to reach a cumulated fluence of $10^{16} \, \mathrm{ 1\,MeV \, n_{eq} / cm^{2}}$, about the one expected in the innermost layers of the tracker during a 10 years lifetime.

\section*{Acknowledgement}

This work was partially funded by: the MINECO, Spanish Government, under grants FPA2013-48308-C2-1-P, FPA2015-69260-C3-2-R, FPA2015-69260-C3-3-R and SEV-2012-0234 (Severo Ochoa excellence programme) and under the Juan de la Cierva programme; the Catalan Government (AGAUR): Grups de Recerca Consolidats (SGR 2014 1177); and the European Union's Horizon 2020 Research and Innovation programme under Grant Agreement no. 654168.

\bibliographystyle{unsrt}

\bibliography{mylibrary}

\end{document}